\documentclass[11pt,twoside]{article}

\usepackage{asp2004}
\usepackage{epsf}
\usepackage{psfig}
\usepackage{lscape}

\begin{document}
\title{The optical and IR properties of southern galactic unclB[e] stars: The case of CD-42$\deg$11721}   
\author{Marcelo Borges Fernandes, Silvia Lorenz Martins}  
\affil{Observat\'{o}rio do Valongo (UFRJ), Ladeira do Pedro Ant\^onio 43, 20080-090 Rio de Janeiro, Brazil}
\author{Michaela Kraus}
\affil{Sterrekundig Instituut, Utrecht University, Princetonplein 5, 3584 CC Utrecht, The Netherlands}
\author{Francisco X. de Ara\'ujo}
\affil{Observat\'{o}rio Nacional-MCT, Rua General Jos\'{e} Cristino 77, 20921-400 S\~{a}o Cristov\~{a}o, Rio de Janeiro, Brazil}

\begin{abstract} 
In the last few years, we had devoted our time to the study of a sample of 
southern unclB[e] stars. For most of them, there are few works in the 
literature, and their physical parameters are still very uncertain. Our 
research was concentrated on the analysis of the optical and IR data, making 
use of high (FEROS) and low (B\&C) resolution spectra, obtained by us at ESO 
(La Silla, Chile) and LNA (Braz\'opolis, Brazil) and using public ISO and IRAS 
spectra. In this work we will present the results for a curious star: 
CD-42$\deg$11721. This object has a doubtful evolutionary stage, being either a 
pre-main sequence star (HAeB[e]) or an evolved star (supergiant B[e] or sgB[e]).
This confusion concerning its nature is caused by the complete absence of 
reliable physical paramenters for this object, especially its distance. Our 
optical investigation could be splitted in two parts, a qualitative study based 
on the identification of the numerous emission lines present in the spectra and 
the classification of their line profiles, which indicate a non-spherically 
symmetric circumstellar environment, and a quantitative analysis of numerous 
forbidden lines, e.g. [O{\sc i}], [O{\sc ii}], [N{\sc ii}] and [S{\sc ii}]. 
Assuming a typical circumstellar scenario for a sgB[e], i.e. a fast, low-density
polar wind and a slow, high-density disc forming equatorial wind, we can 
reproduce very well the line luminosities of the forbidden lines. From this 
analysis, we can determine the mass loss rate of the star lying in the range 
from $\dot{M}\simeq (4.4\pm 0.8)\times 10^{-6}\,{\rm M}_{\odot}{\rm yr}^{-1}$ 
to $\dot{M}\simeq (2.2\pm 0.4)\times 10^{-5}\,{\rm M}_{\odot}{\rm yr}^{-1}$, 
depending on the considered redenning. In addition, our IR study could also be 
splitted in two parts: the identification of several features in the SWS ISO 
spectrum, and the modeling of the SED of CD-42$\deg$11721. 
The first part shows the presence of many unindentified features and 
specially the presence of a mixed chemistry, i.e. C- and O-rich dust in the 
same circumstellar medium. The presence of specially C-rich dust could in 
principle favour a young nature, however this is not so clear, since also other 
evolved stars like e.g. LBVs, show this kind of dust. We have tried to model 
the SED, by using a numerical code written by us which considers a spherical 
circumstellar scenario. Although the answer concerning the evolutionary stage
of CD-42$\deg$11721 is still not very clear, we believe that our analysis will 
improve the discussion about the nature of this curious star.

\end{abstract}


\section{Introduction}   

Unclassified B[e] (unclB[e]) stars cover at least $50\,\%$  
of all galactic stars with the B[e] phenomenon known to date, and their 
number is increasing steadily (see e.g. Miroshnichenko, this volume). 
These stars are concentrated towards the galactic plane, suffering 
severely of a mostly unknown interstellar extinction. Consequently, their 
distances are poorly known, avoiding a proper and reliable determination of
their physical parameters, like effective temperatures, luminosities, and 
abundances, and consequently the comprehension of their nature and evolutionary
stage is rather poor.

\begin{figure}[th]
\plotone{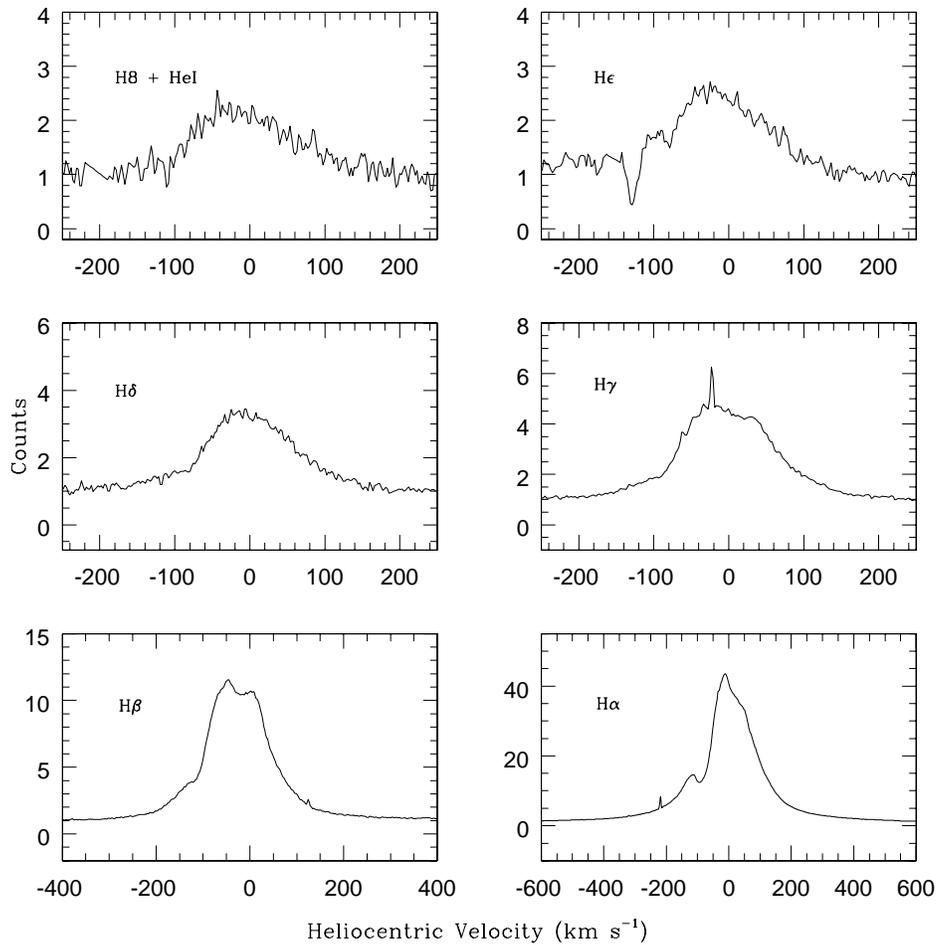}
 \caption{Balmer lines taken from the CD-42$\deg$11721 high resolution (FEROS)
spectrum. The lines are in emission. H$\alpha$ and H$\beta$ show
double peaked profiles, while the higher Balmer lines are single peaked. The 
absorption line blended with H$\epsilon$ is the
interstellar Ca\,{\sc ii} 3968 \AA \ line. Due to the choosen scale
the total extension of the H$\alpha$ and H$\beta$ wings is not clearly visible.}
\end{figure}

In this work, we will present the results obtained for the curious star
CD-42$\deg$11721, based on (i) new observational data obtained by us and using
public IRAS and ISO databases and (ii) by modeling the line luminosities of
the forbidden emisison lines and the dust emission.

\section{Observations}

Intending to improve the discussion concerning the nature of southern galactic 
unclB[e] stars that are very poorly studied, including CD-42$\deg$11721, we 
obtained low and high resolution optical spectra during different observing 
missions at the European Southern Observatory (ESO - La Silla, Chile) and at the
Laborat\'orio Nacional de Astrof\'isica (LNA - Braz\'oplis, Brazil). The high 
resolution spectra (R = 48000) were obtained at the ESO 1.52-m telescope using 
the Fiber-fed Extended Range Optical Spectrograph (FEROS) and 
have a wavelength coverage from 3600\,\AA \ to 9200\,\AA. The low resolution 
spectra ($\sim 4.6$\,\AA) were obtained at the same telescope and at the LNA 
1.6-m telescope using the Boller \& Chivens spectrograph, covering a wavelength 
region from 3800 to 8700\,\AA. 

We are also using public IR spectra taken with the Short Wavelength 
Spectrometer (SWS) on the Infrared Space Observatory (ISO) and taken with the 
Infrared Astronomical Satellite (IRAS). The IRAS Low Resolution Spectra (IRAS 
LRS) cover a small wavelength range, from 7.5 to 22\,$\mu$m, with a spectral 
resolution of about 20-60. On the other hand, the SWS01-ISO spectrum has a 
higher resolution of $\sim$ 1500 and a wavelength coverage from 2 to 50\,$\mu$m. 

\section{The case of CD-42$\deg$11721}   

CD-42$\deg$11721 (V921 Sco, Hen 3-1300, IRAS 16555-4237) was observed for the 
first time by \citet{Merrill}. \citet{deWinter} and 
\citet{Henning} classified it as an HAeBe star, based on the presence of 
a nebulosity, possible spectral and photometric variations, and some further
spectral characteristics, specially in the IR. \citet{Voors} has suggested that 
CD-42$\deg$11721 might be a young star but not a pre-main sequence star because 
of its high temperature and luminosity. On the other hand, a possible nebular 
overabundance of N/O in addition to its high luminosity, as well as some 
spectral similarities to other B[e] supergiants induced \citet{Hutsemekers}
to consider CD-42$\deg$11721 as an evolved object. This is why 
\citet{Lamers} have included it in the list of objects presenting 
the B[e] phenomenon but with an unclear nature, the so-called unclassified 
B[e] stars.

The confusion concerning the evolutionary stage of CD-42$\deg$11721 is strictly 
linked to the absence of any reliable information about its physical parameters.
Since the spectrum does not show any photospheric lines, the spectral type 
ranges from B0 to Aep (depending on the method used) and consequently the 
T$_{\rm eff}$ is lying between 31600\,K and 12300\,K, respectively 
\citep{Hillenbrand,Cidale}. Its luminosity is not well 
known either, ranging between $1.9 < \log(L/L_{\sun}) < 4.9$ 
\citep{McGregor,Shore,Hillenbrand}, not only because of the
uncertain effective temperature, but mainly due to the unclear distance, 
ranging from 136\,pc to 2.6\,kpc \citep{Shore,deWinter,Hillenbrand,Elia},
and the uncertain extinction being either 4.2 or 7.1 mag.

\subsection{Our optical analysis}   

Using our FEROS spectrum (that has a higher resolution 
than any other previously published for this star) we could make a qualitative 
analysis, by identifying the lines and describing their profiles. 

The spectrum of CD-42$\deg$11721 is dominated by emission lines and some (few) 
interstellar absorption lines. We could confirm the presence of the B[e] 
phenomenon and see that Fe\,{\sc ii} is by far the ion with the largest number 
of lines. Other highlights are the Balmer lines presenting very intense 
emission (Fig. 1). We could also identify several forbidden lines of 
[O\,{\sc i}], [O\,{\sc ii}], [S\,{\sc ii}], [N\,{\sc ii}], [Fe\,{\sc ii}], 
[Ni\,{\sc ii}] and [Cr\,{\sc ii}].

Using the Boller \& Chivens spectrum, we could derive the luminosities
of some forbidden lines, e.g., [O\,{\sc i}] ($\lambda$$\lambda$ 5577, 6300,
6363), [O\,{\sc ii}] ($\lambda$$\lambda$ 7318, 7330), [N\,{\sc ii}]
($\lambda$$\lambda$ 5754, 6548, 6584) and [S\,{\sc ii}] ($\lambda$$\lambda$
4068, 4076, 6716, 6731), assuming a mean distance of 1.15 kpc and adopting
the two different values of A$_V$ (4.2 and 7.1).

Next, we calculated the line luminosities of these forbidden lines, considering
either a spherical scenario (to prove or disprove the necessity of a disc)
or a two-component wind scenario. By modeling self-consistently the observed 
line luminosities it is possible to fix the mass loss rate of the star
\citep[see e.g.][]{Ketal2005}.   
Our best results were obtained with the two-component wind scenario, i.e. 
assuming a polar wind plus an equatorial disc (with an open angle of 
$\sim 25\deg$). We assumed an effective temperature for CD-42$\deg$11721 of 
15\,000\,K, and terminal wind temperatures of 12\,000\,K for the polar wind 
and 8\,000\,K for the equatorial disc. The results for the [N\,{\sc ii}] lines, 
which are produced in the polar wind only, are shown exemplarily in Fig. 2.

\begin{figure}[th]
\plotone{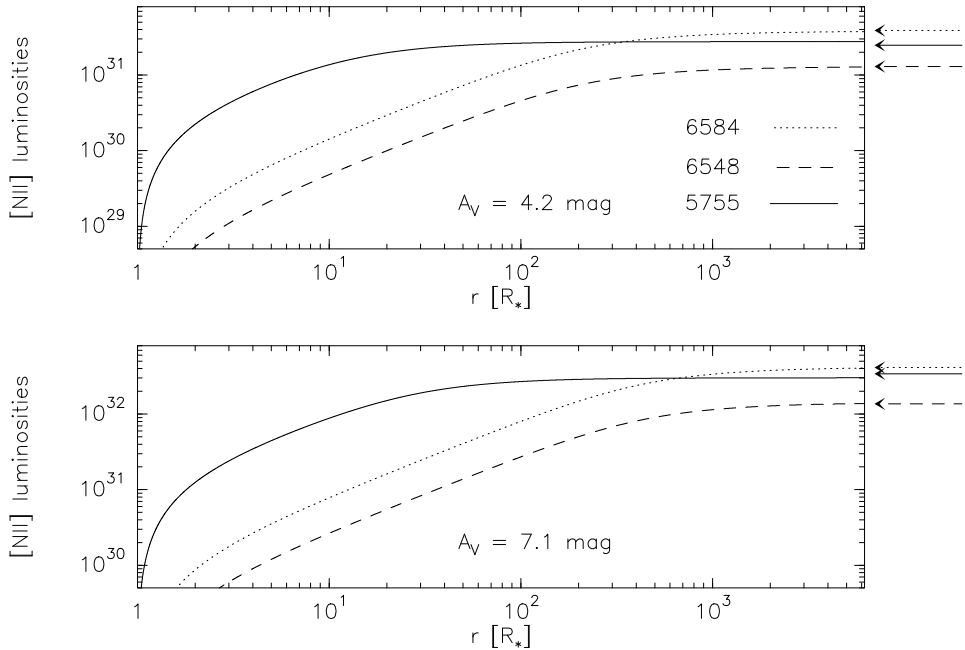}
\caption{Model results for the [N{\sc ii}] lines considering the two different 
reddenings. These lines are produced in the polar wind only. 
The arrows on the right side indicate the observed value for each emission line
(same line style). The outer edge of the modeling is set by the slit width 
used and the adopted distance to CD-42$\deg$11721.}
\end{figure}

The derived mass loss rates, accounting for the two different extinction
values of 4.2 and 7.1 mag, are $4.4\pm 0.8\times 10^{-6}\,{\rm 
M}_{\odot}{\rm yr}^{-1}$ and $2.2\pm 0.4\times 10^{-5}\,{\rm M}_{\odot}{\rm 
yr}^{-1}$, respectively. Interestingly, when modeling the line luminosities
it turned out that nitrogen has to be overabundant compared to solar, while
all other elements could be modeled with solar abundances. This N overabundance
and the high mass loss rates found from the modeling  
favour a supergiant nature (Borges Fernandes et al. 2005A, in preparation).

\subsection{Our IR analysis} 

We have also made a description of the CD-42$\deg$11721 infrared spectrum based 
on IRAS LRS and especially SWS01 ISO public data. We could identify several new 
features that were not cited previously in the literature (Borges Fernandes et 
al. 2005B, in preparation). The presence of the many H recombination lines 
and the forbidden lines is remarkable, and especially the so-called "dual-dust" 
chemistry \citep{Waters}, characterized by the simultaneous presence 
of solid state bands of C-rich and O-rich dust being produced in the same 
environment. The C-rich material is represented by PAH emission bands 
\citep{Jourdain} and the O-rich dust by some features caused 
by crystalline silicates \citep{Voors}.

\subsection{Modeling of the SED} 

Using photometric data from the literature, we could obtain the spectral energy 
distribution (SED) of CD-42$\deg$11721. It is double peaked, similar to those 
called ``group I" of HAeBe stars \citep{Meeus}. We have tried to model 
this SED, considering different physical parameters (including those 
obtained by our optical study), using a code that treats the radiative transfer 
in a sherical envelope making use of the Monte Carlo method \citep{Silvia}.
As can be seen in the Fig. 3, we could not model properly this 
SED, indicating that a non-spherical model, most plausibly a disc scenario,
is indeed necessary to account for the observed SED, in agreement with 
our optical analysis. A study involving a non-spherical dust distribution
is currently under investigation (Borges Fernandes et al. 2005B, in 
preparation).

\begin{figure}[th]
\plotone{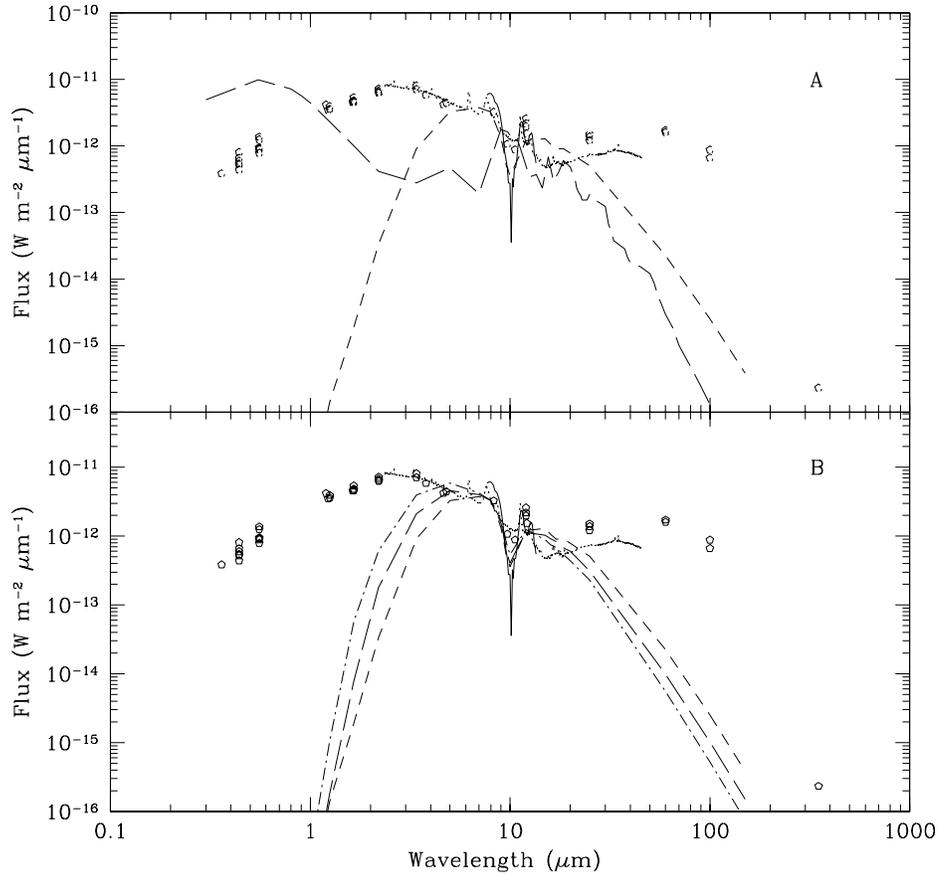}
\caption{Different spherical models trying unsuccessfully to reproduce the SED
of CD-42$\deg$11721.} 
\end{figure}

\subsection{Conclusions}

CD-42$\deg$11721 is a very curious star whose evolutionary stage is 
still unclear. Our analysis resulted in arguments for as well as against
the classification as either a young or an evolved object:

In favour of a young HAeBe nature speak its IR characteristics, specially its 
double peaked SED and the position of this object in color-color diagrams.
However, we can not discard a possible contamination of the photometric data,
since CD-42$\deg$11721 is lying in a crowded region, hosting several stars
as well as a cloud which might be of interstellar nature. 

In favour of a supergiant nature speaks our optical analysis which shows 
excellent results considering a supergiant two-component wind scenario, and 
which, in addition, indicated the necessity of an N overabundance.
Against this scenario speaks, however, the dual-dust chemistry, specially
the presence of C-rich dust which is difficult to explain with a supergiant
scenario whose spectrum is dominated by H recombination lines.
We clearly need more multiwavelength and high spatial resolution observations 
of CD-42$\deg$11721 and its 
close-by circumstellar medium for a better determination of its nature.

\section{Summary}	

In this work, we are showing clearly how hard the life is for those people that 
are studying unclB[e] stars. However, the future is very promising since new 
observations using different techniques, like high resolution spectroscopy, 
spectropolarimetry and interferometry in different spectral regions, will allow 
us to understand better on the one hand the nature of these fascinating objects,
and on the other hand some important stellar evolutionary phases that are still 
poorly known and investigated.

\acknowledgements 

M.B.F. acknowledges financial support from \emph{CNPq} (Post-doc position - 
150170/2004-1), \emph{Utrecht University}, \emph{LKBF} and \emph{NOVA} 
foundations. M.B.F. also acknowledges L.B.F.M. Waters and H.J.G.L.M. Lamers for 
pleasant and fruitful discussions about the nature of CD-42$\deg$11721 on 
Vlieland, and specially M. Kraus for the excellent conference organization and 
also for all the support during my stay. M.K. acknowledges financial support 
from the Nederlandse Organisatie voor Wetenschappelijk Onderzoek grant 
No.\,614.000.310.

\end{document}